\documentstyle[prb,aps,twocolumn,floats]{revtex}

\begin{document}

\title {\bf Weak localisation, interaction effects and the
metallic phase in p-SiGe}

\author{P.T. Coleridge, A.S. Sachrajda and P. Zawadzki \\ }

\address
{Institute for Microstructural Sciences, National Research
Council, Ottawa,
Ontario, K1A OR6, Canada\\ }

\date{21 December 2001}

\maketitle

\begin{abstract}

Magnetoresistance results are presented for p-SiGe samples on
the metallic side of the B=0 metal-insulator transition. It was
possible to separate the weak localisation and Zeeman
interaction effects but the results could not be explained
quantitatively within the framework of standard theories for
quantum corrections of a weakly interacting 2-dimensional
system. Analysis using a theory for interaction corrections at
intermediate temperatures, recently proposed by Zala, Narozhny
and Aleiner, provided values of the Fermi liquid parameter
$F_0^{\sigma}$ of order -0.5. Similar values also explain the
linear increase of resistance with temperature characteristic
of the metallic phase.
 
\vspace*{2.0cm}


\end{abstract}


\subsection{Introduction}

     Recent experiments indicating the existence of a
metallic state and a metal-insulator transition (MIT) in two-
dimensional (2-D) semiconductor systems \cite{} continue to
attract attention \cite{abrahams00}. There is, as yet, no
general consensus about the origin of the metallic behaviour
and it remains a controversial topic.  The metallic behaviour
appears in strongly interacting systems where $r_s$ (the ratio
of the interaction energy to the kinetic energy) is large,
typically 5 to 20 and the Coulomb interaction energy is by far
the largest energy in the problem. While the existence of a MIT
in 2-dimensional systems contradicts the well established one
parameter scaling theory for non-interacting systems
\cite{abrahams79}, it is not {\it a priori} forbidden for
strong interactions \cite{dobr97} and the observation of good
scaling behaviour \cite{simonian97}, with symmetry about the
critical density, supports the view that this is a genuine
transition, driven by the interactions. This is also
consistent with predictions of renormalisation group (RG)
theories
\cite{abrahams99,finkelstein84,castellani84,castellani98} that
a low temperature metallic phase can exist when interactions
and disorder are both important. 

     An alternative view \cite{altshuler99} is that there is no
transition, just a cross-over from weakly localised to strongly
localised behaviour, and that the strong interactions do not
significantly modify the basic Fermi liquid character. Large
values of $r_s$ are almost inevitably associated with low
densities and Fermi energies ($k_B T_F$) that are not much
larger than the measuring temperatures so effects associated
with the small Fermi degeneracy are likely to be significant.
It is argued that these can fully account for the metallic-like
increase in resistivity with temperature and that weak
localisation and interaction effect corrections are present but
concealed by the larger semi-classical effects.

     Experimental support for this second viewpoint has
recently been presented based on results obtained in p-GaAs
\cite{simmons00} and p-SiGe \cite{senz00b}. In both cases the
B=0 resistivity shows no direct evidence of a lnT dependence,
the standard signature of both weak localisation and
interaction effects but the magnetoresistance has the
characteristic negative peak associated with the destruction,
by dephasing, of weak localisation. Also the Hall coefficient
shows a lnT dependence, interpreted as evidence for interaction
corrections. The magnitude of both these terms is consistent
with the standard predictions and it is argued the behaviour is
that of an entirely conventional Fermi liquid, despite the
large values of $r_s$.

     Similar measurements, in p-SiGe samples, are presented
here \cite{ptc99}. The experimental data is generally
consistent with results obtained in other samples but is more
detailed and is interpreted somewhat differently.  It is
suggested that the behaviour cannot be explained using the
standard theories for weakly interacting Fermi liquid systems.
Not only are the parameters characterising the interaction too
large to justify the assumption of weak interactions but also
there are internal inconsistencies in the analysis. At low
temperatures the saturation of resistance is generally
consistent with the predictions of renormalisation group
theories and at higher, intermediate, temperatures the results
can be understood within the framework of a new theory for
interaction effects
\cite{zala01a}. In both cases a large amplitude of spin
fluctuations plays an essential role.

\subsection{Theory: weak interactions}

     Because the data presented below is discussed,
initially, in terms of the quantum corrections for weakly
interacting systems it is convenient to summarise the
standard theoretical treatments
\cite{altshuler85,fukuyama85,lee85} which have been used,
historically, to analyse experiments. They start with a semi-
classical description of the transport, where the diffusion of
electrons at the Fermi level is considered to obey
classical dynamics and quantum effects are introduced as small
corrections. Drude-Boltzmann theory gives the zero field
conductivity as 

\begin{equation}
     \sigma_0  =  n_s e^2 \tau/m^{\ast}
               = \frac{g_s} {2} k_f l \frac{e^2} {h}
\label{eq1}
\end{equation}
where $n_s$ is the carrier density, $\tau$ the transport
lifetime, m$^{\ast}$ the effective mass, $g_s$ the spin
degeneracy, $k_f$ the Fermi wavevector and {\it l} the mean
free path. In a magnetic field the conductivity components are
given by

\begin{equation}
\sigma_{xx}(B) = \frac{\sigma_0} {(1  +  \mu^2 B^2 )};
\;\;\;\;    \sigma_{xy}(B) = \mu B \sigma_{xx}(B)
\label{eq2}
\end{equation}
where the mobility $\mu$ = $e\tau / m^{\ast}$. Quantum
interference introduces a weak localisation correction to this
conductivity with a logarithmic temperature dependence 

\begin{equation}
\Delta \sigma_{xx}^{wl}(T) =  
     \alpha p  (e^2/\pi h) \ln (k_B T \tau/ \hbar)
\label{eq3x}
\end{equation}
where it is assumed the phase breaking time $\tau_{\phi}$
varies as $T^{-p}$. The amplitude $\alpha$ is expected to be 1
for normal scattering (-0.5 for pure spin-orbit scattering and
0 for spin scattering).

A lnT dependence also results from the Coulomb interaction
effect  

\begin{equation}
\Delta \sigma_{xx}^{ee}(T) =  (1- \frac{ \textstyle 3
F^{\ast}} { \textstyle 4}) (e^2/ \pi h) \ln (k_B T \tau/ \hbar)
.
\label{eq4x}
\end{equation}
Two processes, with opposite signs, contribute here. A
Hartree term (singlet channel), involving only small momentum
transfers, and an exchange term (triplet channel) involving
$F^{\ast}$, the Fermi surface average of the screened Coulomb
interaction. Increasing $F^{\ast}$ implies an increased
tendency towards delocalising behaviour. For Thomas-Fermi
screening $F^{\ast} < 1$ but it may, in principle, be larger.
In this case a weakly interacting theory is inappropriate and
should be replaced by more sophisticated approaches but with
trends that are probably still given correctly by eqn.4. Large
values of $F^{\ast}$ imply negative coefficients for the
interaction term which may even overcome the weak localisation
term and result in a total negative, or delocalising, lnT
dependence. Values of $F^{\ast}$ larger than one are frequently
obtained when fitting experimental data
\cite{bishop82,burdis88}.

     Application of a magnetic field allows the weak
localisation and interaction terms to be separated.  At low
fields dephasing of the weak localisation term gives a
characteristic , negative magnetoresistance \cite{hikami80},

\begin{equation}
  \Delta \sigma_1 (B) =  \alpha \frac {e^2} {\pi h}
     [ \Psi ( \frac{1}{2}+\frac{\tau_B}{2 \tau_{\phi}})
  - \Psi ( \frac{1}{2}+\frac{\tau_B}{2 \tau})  + 
     \ln ( \frac{\tau_{\phi}}{\tau})]
\label{eq5x}
\end{equation}
where $\Psi$ is the digamma function and $\tau_B =
\hbar/2eDB$ with D (the diffusion constant) = $v_F^2 \tau /2$.
In terms of the variable $h=2\tau_{\phi}/\tau_B$
(proportional to B and not to be confused with the Planck
constant) this function varies as $h^2/24$ for small {\it h}
and as $\ln(h)$ for $1 \ll h  \ll \tau_{\phi}/\tau$.  Fitting
to the characteristic shape, particularly at small B, allows an
experimental determination of $\alpha$ and  $\tau_{\phi}$.
Although commonly used to analyse experimental data eqn.5 is
known to be in error for large fields, when the diffusion limit
is no longer valid. In this case, however, it has been shown
\cite{minkov00} that the equation still gives a good fit to the
data but with smaller values of $\alpha$ (typically 0.5 -0.7)
and fitted values of $\tau_{\phi}$ that are in error by
10-20\%.

     Magnetic fields corresponding to $h \sim 1$,
sufficient to dephase the weak localisation, are too small to
significantly affect the interaction term  but higher fields,
large enough to produce a Zeeman spin-splitting, modify the
triplet part of this term.  This results in a positive
magnetoresistance given by

\begin{equation}
  \Delta \sigma_2 (B)  =  - ( e^2 / \pi h) ( F^{\ast}/ 2) G(b)
\label{eq6x}
\end{equation}
where $b = g^{\ast}\mu_B B/ k_B T$ with $g^{\ast}$ the g-factor
of the spins. The function G(b) is known: for small b  $G(b) = 
0.084 b^2$, for large b it varies as $\ln(b/1.3)$,  and it can
be calculated in the intervening region
\cite{burdis88}.

In addition to these terms there is also the classical
magnetoconductance (eqn.2). For small B, this gives a quadratic
correction term

\begin{equation}
     \Delta \sigma_{xx}(B) \approx   - \sigma_0  \mu^2 B^2.  
\label{eq7x}
\end{equation}

     For the Hall conductivity $\sigma_{xy}$ the weak
localisation corrections appear a factor of two larger
\cite{altshuler85,fukuyama85,lee85} but there are no
interaction corrections. Therefore, when the Hall coefficient
($R_H =  \rho_{xy} / B$ ) is obtained by inverting the
conductivity tensor the weak localisation terms cancel and 

\begin{equation}
     \Delta R_H / R_H    
     =  -2  \Delta \sigma_{xx}^{ee} /\sigma_{xx}.
\label{eq8x}
\end{equation}
The Hall coefficient is therefore expected to have a
logarithmic temperature dependence (given by eqns.4 and 8) and
a field dependence (eqn.6) but should not include any weak
localisation corrections.

     If, in low fields,  $\rho_{xx}$  rather than
$\sigma_{xx}$ data is analysed the lnT weak localisation and
interaction effect terms appear additively in the same way but
the quadratic classical term (eqn.7) is automatically cancelled
by the $\sigma_{xy}$ contribution. At higher fields, when $\mu
B$ becomes significant, the admixture of $\sigma_{xx}$ and
$\sigma_{xy}$ contributions introduces an additional term 
\cite{houghton82}, $\Delta \rho_{xx} = (\mu^2 B^{2}-1) \Delta
\sigma_{xx}^{ee} / \sigma_0^2$ ie a parabolic negative
magnetoresistance with a temperature dependence given by eqn.4.

\subsection{Samples}

     The samples used in this investigation are from a set of
p-type modulation doped strained Si-Ge quantum wells which
exhibit MIT behaviour \cite{ptc97}. Grown using a low
temperature UHV-CVD process they have a Si buffer layer, a 40nm
Si$_{.88}$Ge$_{.12}$ quantum well, a spacer layer (of variable
thickness), a boron doped layer and a thin Si cap. The holes
reside in an approximately triangular SiGe quantum well,
produced by the asymmetric doping, which is strained because it
is lattice matched to the Si substrate. It has been established
that they are in almost pure  $|M_J|$ = 3/2   states, well
decoupled from other hole states by strain and confinement
\cite{hensel63,people85}. Most of the data presented here comes
either from Sample A,  with a density ($p_s$) of
5.7$\times$10$^{11}$cm$^{-2}$ which is deep in the metallic
phase or from sample B, with a density of
1.2$\times$10$^{11}$cm$^{-2}$, which is close to the critical
density for the MIT ( approximately
1.0$\times$10$^{11}$cm$^{-2}$). Values of $r_s$ (defined as
$1/(\pi p_s)^{1/2} a^{\ast}$, where $a^{\ast}$ is the effective
Bohr radius using the semiconductor dielectric constant) are 
4 and 6, respectively, for the two samples.  

     Effective mass values are known only approximately in this
system. Measurements from the temperature dependence of the
Shubnikov-de Haas oscillations \cite{ptc96} gave values of
0.30$_5$m$_0$ for sample A and approximately 0.23m$_0$ for
sample B. These are significantly larger than the band mass of
0.20m$_0$ \cite{hasegawa63}. It is possible this reflects a
breakdown of the standard Lifshitz-Kosevitch expression in a
system with strong quantum corrections. High field cyclotron
resonance measurements \cite{song95,wong95} have given values
as low as 0.18m$_0$.

     The g-factor ($g^{\ast}$) in p-SiGe is also not well known
but is of order 4 in perpendicular fields \cite{glaser90}. In
sample A the low field Shubnikov-de Haas (SdH) oscillations
appear initially at odd filling factors \cite{ptc96},
corresponding to a Zeeman splitting of the Landau levels
sufficiently large that $g^{\ast}m^{\ast}/m_0 > 1$. With 
$m^{\ast}/m_0 $ = 0.25 - 0.30 $g^{\ast}$ is therefore of order
4. In sample B the SdH oscillations appear initially at even
filling factors so $g^{\ast}m^{\ast}/m_0 < 1$ . Using
m$^{\ast}$ = 0.23m$_0$  and taking $g^{\ast}m^{\ast}/m_0$ to be
0.8$\pm$ 0.2 gives $g^{\ast} = 3.6 \pm 1$.

\subsection{Experimental procedures}

     Measurements were made using a DC bridge \cite{avs}, with
current reversal at 15Hz. Two cryostats were used, a dilution
refrigerator giving temperatures below 100mK and a sorption
pumped He3 system that could achieve sample
temperatures down to 270mK. In both cases thermometry was in a
field free region, with the sample cooled predominantly by the
copper leads. Measurements using a calibrated, field
insensitive, thermometer in place of the sample indicated
temperature gradients of less than 5mK.

     Resistance measurements, especially in sample B, showed an
unusual sensitivity to measuring current at the lowest
temperatures. Currents as low as 1nA, corresponding to
voltages of only a few microvolts, produced detectable
metallic-like  non-linearities in the I-V characteristics and
made accurate low temperature measurements of the zero field
resistivity difficult. In addition, sweeping the magnetic field
caused abnormally large heating (and cooling) effects. These
phenomena are not well understood but are attributed, at least
in part, to the proximity of the MIT. Very slow field sweeps
were used with frequent checks (making
corrections if necessary) to ensure that the measured
resistances were the same in static and swept fields. In the
He3 system measurements in magnetic fields were also
complicated by a superconducting solenoid that had an
unexpectedly large degree of trapped flux. The
associated hysteresis was, however, found to be reproducible
and a calibration procedure could be established that
resulted in field errors, around B=0, estimated to be less than
1 mT.

\subsection{Experimental results} 

     Magnetoresistance data from sample A is shown in figure 1.
At zero field the temperature dependence of the resistance is
characterised by an approximately constant low temperature
region followed by a monotonic, metallic-like, increase with
temperature. This is characteristic of samples showing a MIT.
For this sample, at 20K, the resistance has increased by a
factor of roughly two. Over the temperature range shown (0.15 -
 2.2K) a logarithmic weak localisation term (eqn.3) would be
expected to give a {\it decrease} of resistivity of order 100
ohms/square, roughly a factor of two larger and in the opposite
sense to the actual experimentally observed change. The well
defined low field negative magnetoresistance, with a width that
increases with temperature, is attributed to weak localisation; 
the positive magnetoresistance at higher fields to the Zeeman
interaction term. 

     Attempts to fit the field dependence to a combination of
eqns. 5 and 6 failed, even when five separate fitting
parameters (including the g-factor) were used. The alternative
approach taken was therefore to fit the data to eqn. 5 at the
lowest fields, using just three parameters: $\tau$ (determined
essentially by the B=0 value of the
resistivity), $\tau_{\phi}$ and an amplitude $\alpha$. The
results of these fits are shown in figure 1. Care was taken to
establish that the parameters were insensitive to the precise
range of B and also that they were not significantly altered
when this range was extended and an additional
quadratic term included. Values of $\tau_{\phi}$,  shown in
figure 2, are similar to those obtained from other
measurements in p-SiGe \cite{emeleus93,cheung94,senz00a}.
Fitted values of $\alpha$ were 0.7 at the lowest temperatures
decreasing to 0.6 at higher T.

      The residues from these fits are shown in fig.3a.
Plotted against B/T (fig. 3b) they collapse onto a single curve
which gives some confidence that the fitting procedure has
successfully separated the weak localisation term from a Zeeman
interaction term which is a function of B/T. This curve is not,
however, given by eqn.6. The solid line in fig. 3b is G(b),
fitted to the curve for larger values of B/T, with $F^{\ast}$
= 2.45 and a g-factor of 6.4. This deviates significantly from
the data at low values of B/T and the value for $g^{\ast}$ is
somewhat larger than expected. We believe this deviation is not
an artefact of the fitting procedure but rather is consistent
with the fact, noted above, that it was impossible to fit
individual data curves to a combination of equations 5 and 6.

     The most important conclusion, despite the only
qualitative agreement between theory and experiment, is the
large value of $F^{\ast}$. This implies that the interactions
are strong and that the standard theory is probably inadequate.

     Expressions for the Zeeman interaction term have also been
obtained within the framework of renormalisation group theory
by Finkel'stein \cite{finkelstein84} for b $\gg$ 1,

\begin{equation}
  \Delta \sigma_2 (B)  =  - ( e^2 / \pi h) \, 2 \, 
  [\frac{1 + \gamma_2}{\gamma_2} \ln(1 + \gamma_2) - 1 ]      
 \ln ( b )
\label{eq9x}
\end{equation}
and by Castellani {\it et al }
\cite{castellani84,castellani98} for b $\ll$ 1  

\begin{equation}
  \Delta \sigma_2 (B)  =  - 0.084 ( e^2 / \pi h) \gamma_2  (1
+ \gamma_2) b^2 .
\label{eq10x}
\end{equation}
Here $\gamma_2$ is a measure of the renormalised interaction
strength for the triplet scattering channel. It reduces to
F$^{\ast}/2$ in the limit of weak interactions.

     The dashed lines in fig. 3b show fits using these two
equations in the high field and low field limits respectively,
using $g^{\ast}$ = 3.6 and $\gamma_2$ = 2.6 in both cases.
Interpolating between the high field and low field expressions
then gives a very reasonable description of the experimental
data. The value of $g^{\ast}$ is perhaps a little small but
this is predominantly determined by the low field quadratic
regime where experimental errors are relatively large.

     It might perhaps be argued that because eqn.5 is invalid
at high fields, in the ballistic regime, the subtraction of the
weak localisation term introduces an error and the residues
should not be attributed solely to a Zeeman term. It has been
shown, however \cite{minkov00}, that this equation provides an
adequate description of the weak localisation in this regime
but with a prefactor $\alpha$ that is less than unity. The
values of $\alpha = 0.6 - 0.7$ determined experimentally are
consistent with this interpretation and imply that the residues
can correctly be identified with the Zeeman term. Further
confirmation is provided by the fact that any spurious
contribution from the weak localisation term would not have the
observed functional dependence on B/T.

     Magnetoresistance data for sample B is shown in figure 4.
Here, again, the zero field increase of resistance with
increasing temperature in the opposite sense to that expected
for weak localisation. Over the temperature range shown the
expected lnT weak localisation term would correspond to a
decrease in resistivity of approximately 2000 ohms/square. The
negative magnetoresistance associated with weak localisation is
less well defined than in sample A, mainly because the ratio
$\tau_{\phi}/\tau$ is significantly smaller. Because of this
and because the Hall coefficient is temperature dependent (see
below) it was not possible to just fit $\rho_{xx}$ and rely on
the Hall term to cancel the classical quadratic term (eqn.7). 
Rather it was necessary to explicitly determine and fit the
conductivity $\sigma_{xx}$. Shown in figure 5 (and obtained by
inverting the measured values of $\rho_{xx}$ and $\rho_{xy}$)
this is dominated by a quadratic dependence on field that is
the sum of the classical term and the low field Zeeman
interaction term (eqns. 6 or 10).

     The rather small negative magnetoresistance associated
with the weak localisation made it impossible to use the
fitting procedure developed for sample A: the equations became
ill-conditioned and gave imprecise and interdependent values of
$\alpha$ and $\tau_{\phi}$. The fit is relatively insensitive
to the exact value of $\alpha$ (see also Senz {\it et al}
\cite{senz00a}) so this was fixed (at 0.65) and just three
adjustable parameters used: $\tau$, $\tau_{\phi}$ and a
quadratic coefficient. As can be seen in figure 5 these gave
very good descriptions of the data provided the fits were
restricted to the field region where the Zeeman interaction
term is expected to vary quadratically with B. Values of
$\tau_{\phi}$ are shown in figure 2.

     Subtracting off the weak localisation and classical
mobility terms should leave the Zeeman interaction term but
there is some uncertainty about exactly how to determine the
classical term.  In practice $\sigma_0$ was obtained at each
temperature by adding the amplitude of the weak localisation
correction, ie $\alpha (e^2/\pi h) \ln (\tau_{\phi}/\tau)$, to
the measured value and then using this, with the known density,
to obtain the mobility.  This procedure was applied self-
consistently so the value of $\tau$ determined from the
mobility was also used in the weak-localisation fit. The Zeeman
interaction term obtained in this way is plotted against B/T in
figure 6.  The errors could be quite large here, for example
the classical mobility correction is approximately equal to the
Zeeman term at the highest temperatures, but the good collapse
onto a single curve suggests they are not, in fact,
significant.

     The dashed line in figure 6 shows a fit of the data to
G(b) assuming a g-factor of 3.6. In contrast to sample A
(figure 4b), the deviation away from quadratic towards
logarithmic behaviour at higher fields appears to be
adequately described by eqn. 6 and corresponds to
$F^{\ast}$ = 1.95.  Alternatively, using eqn. 10 in the
quadratic region,  $\gamma_2$ = 0.6. These values are
significantly smaller than in sample A. Taking eqn. 4 at face-
value, and assuming a weak localisation term is present, the
value of $F^{\ast}$ = 1.95 is close to the cross-over between
a net localising or delocalising lnT dependence.

     According to the standard approach an independent measure
of $F^{\ast}$ can be obtained from the temperature dependence
of the Hall coefficient. This should have a lnT dependence
given by eqns.4 and 8, ie be proportional to
$(1-3F^{\ast}/4)$.  Measurements of $R_H$ at very low fields
are complicated any small magnetic field error. Cubic fits to
$\rho_{xy}$ were made in this region and matched onto direct
measurements at higher fields \cite{footnote1}. Results are
shown in figures 7 and 8. 

     In sample A the very small, non-monotonic, temperature
dependence around B=0 is of the same magnitude as the estimated
experimental errors. Over the temperature range shown $R_H$
deviates by less than 2\% from 1090 ohms/tesla, the value
corresponding to a density of 5.7$\times$10$^{11}$cm$^{-2}$
determined from the periodicity of the Shubnikov-de Haas (S-dH)
oscillations. Taken at face value this implies $F^{\ast}
\approx 1.33$, roughly a factor two smaller than the value
determined above from the Zeeman interaction term.  At higher
fields a significant field and temperature dependence develops.
Unlike the data shown in figure 3b this is not a good function
of B/T and is also a factor of roughly two smaller than
expected for $F^{\ast} \sim$ 1.3. It should be noted, in
contrast to measurements recently reported in p-GaAs
\cite{proskuryakov01}, that these corrections to the Hall
coefficient are not reflected in the Shubnikov-de Haas
oscillations. The period of these oscillations, and of the
corresponding oscillations in $\rho_{xy}$ that can be seen in
figure 7, has no significant temperature dependence.

     In sample B (figure 8) the Hall coefficient at B=0 shows
a clear lnT dependence (see inset) and is significantly
increased over the value of 5400 ohm/tesla that corresponds to
the density of 1.16$\times$10$^{11}$cm$^{-2}$ obtained from the
periodicity of the Shubnikov-de Haas oscillations . Although
less evident than in sample A, the period of these is again,
unaffected by the enhanced Hall coefficient. Using eqns. 4 and
8 the slope of the lnT dependence shown in the inset ($\Delta
R_H/R_H = -.04 \ln(T)$) corresponds to $F^{\ast}$ = 1.1.  The
approximately quadratic increase of $R_H$ with magnetic field
is again, a poor function of B/T and less than half the
expected value.  

     Within RG theory eqn. 4 is modified to
\cite{finkelstein84} 

\begin{equation}
\Delta \sigma_{xx}^{ee}(T) =  [4-3\frac{1+\gamma_2}{\gamma_2}
\ln(1+\gamma_2)] (e^2/ \pi h) \ln (k_B T \tau/ \hbar) .
\label{eq11x}
\end{equation} 
Interpreted using eqn.8 this implies $\gamma_2$ = 0.65.

     For temperatures such that $k_B T \tau/\hbar \sim 1$ the
screening function, and therefore the Drude scattering time,
becomes temperature dependent. The corresponding temperature
dependent conductivity has been calculated by Gold and
Dolgopolov \cite{gold85} as

\begin{equation}
     \sigma(T) = \sigma(0) [1 - C_p (T/T_F)]. 
\label{eq12x}
\end{equation}
Similar results have also been obtained numerically by Das
Sarma \cite{dassarma86}. 

     Senz {\it et al} \cite{senz00b} on the basis of the
analysis of measurements in other p-SiGe samples, have
suggested that this can account for the absence of a weak
localisation lnT term in the zero field conductivity. They
argue that such a term is in fact present but when combined
with the linear temperature dependent screening result gives an
{\it apparent} low temperature saturation of resistivity. A
similiar analysis is presented here: for samples A and B and
also for sample C (which has a density very similar to sample
B but a significantly larger peak mobility) and sample D which
has a very similar density to sample A but where the
measurements extend to a much lower temperature.

     The solid points in figure 9 show the measured
conductivities in these samples, plotted against $T/T_F$. Open
points show the result of subtracting a putative weak-
localisation correction $\alpha p (e^2/\pi h) \ln(k_B T
\tau/\hbar)$ and the lines are linear fits to the high
temperature data.  At higher temperatures the modified
results, like the raw data, all exhibit the predominantly
linear dependence predicted by eqn. 12. The coefficients C$_p$
are, however, larger than expected. With impurity and interface
roughness scattering dominating, C$_p$ for samples A and D is
expected \cite{plews97} to be 1.2 - 1.5,  compared with
experimental values of 2.5 and for samples B and C the values
of 5.0 and 6.1 shown are two to three times the expected value
of about 2. Also, in the adjusted data, there are significant
deviations from a linear dependence at the lowest temperatures.
The upturn there reflects directly the lnT divergence of the
weak localisation term subtracted from approximately constant
experimental data.

     Although in qualitative agreement with the general picture
proposed by Senz {\it et al} these results do not support their
detailed conclusions.  The data presented here cannot be
explained as just the sum of a lnT weak localisation term and
a linear temperature dependent screening term. There is either
a real saturation of the resistivity at low temperatures or if
the weak localisation lnT term is present it must be cancelled
by another lnT term of the opposite sign. Also the magnitude of
the linear term is larger than expected. 

\subsection{Discussion}

     When the data presented above is analysed within the
framework of the standard theory for weakly interacting
systems many of the {\it trends} can be understood but there
are quantitative disagreements. For much of the data the
corrections are not small, as is assumed in the theory, but are
comparable to the classical Drude term. The low field
negative magnetoresistance is consistent with weak localisation
but the expected lnT dependence appears to be replaced by a
positive, approximately linear, temperature dependence. This
linear increase transforms smoothly into insulating behaviour
at the critical density, which, as in Si-MOSFETs can be
described by a scaling expression \cite{ptc97,senz99}.

     RG theories predict that in the presence of disorder the
Fermi liquid parameter $\gamma_2$ measuring the strength of the
spin fluctuations becomes temperature dependent. This leads to
a decrease in resistance with decreasing temperature that
eventually saturates at a finite value. The absence of any lnT
dependence might possibly be explained if both the weak
localisation and interaction terms were to be renormalised in
this way. The magnetoresistance of the weak localisation,
however, shows little evidence of such behaviour and appears to
be entirely conventional. For the values of $\tau_{\phi}$ shown
in figure 2 the parameter $k_B T \tau/\hbar$ varies between
about 0.02 and 0.3 in which case the dephasing should be
determined by inelastic hole-hole scattering with small
momentum transfer (sometimes known as Nyquist dephasing) and
$\tau_{\phi}$ is given by
\cite{altshuler85,senz00a,brunthaler00,altshuler82} 

\begin{equation}
     \tau_{\phi} = \hbar f(g)/k_B T  
\label{eq13x}
\end{equation}
where $f(g)$, with the conductance {\it g} in units of
$e^2/h$, is $g/ \ln(g/2)$ for $g \gg 1$ but of order {\it g} as
{\it g} approaches 1. These values are shown as lines with
$f(g) = 7.4$ for sample A (where $g\approx15$) and 2.6 for
sample B (where $g\approx2.6$). Experimental values of
$\tau_{\phi}$ are four to five times smaller.  A similar
discrepancy is also seen in Si-MOSFETs \cite{brunthaler00}, in
p-GaAs \cite{simmons00,proskuryakov01} and in other p-SiGe
experiments \cite{emeleus93,cheung94,senz00a}: it is not clear
whether this discrepancy is general or confined to systems with
large values of $r_s$.

     Interaction effects are clearly large. While the Zeeman
effect data in sample A does not have the expected functional
form $F^{\ast}$ is at least as large as 2.5. For sample B,
where there is a better fit to the theory,  $F^{\ast}$ is also
large. From the temperature dependence of the Hall coefficient
$F^{\ast} \sim 1.2$ in both samples, roughly a factor two
smaller than the values deduced from the Zeeman interaction
effect.

     If the standard theory is assumed to be valid for these 
large values of $F^{\ast}$ the absence of a lnT term in
$\sigma_{xx}$ might perhaps be attributed to a cancellation
between a positive coefficient for the weak localisation
correction and negative coefficient for the interaction
correction with the sum, $(\alpha p + 1-\frac{3}{4} F^{\ast})$, 
vanishing. While this might happen fortuitously in one sample
it seems unlikely for two samples where, according to the
Zeeman term, the values of $F^{\ast}$ are quite different. It
would not explain why, in sample A, the lnT term is absent in
both $\sigma_{xx}$, where the weak localisation is included,
and $R_H$, where it is excluded. Furthermore, in sample B the
measured lnT term in $R_H$ is of the wrong sign to cancel a
weak localisation term.

    In these samples therefore, with large values of $r_s$, the
standard weakly interacting theory provides an inconsistent and
inadequate description of the data. The coulomb interactions
are large, probably large enough to overcome weak localisation
effects and lead to a net delocalising or metallic-like
temperature dependence.  Near the critical density for the MIT
the contributions to the conductivity from disorder, weak
localisation, coulomb interactions and temperature dependent
screening effects are all important and all comparable to the
classical Drude term so it seems likely that a satisfactory
theoretical description of the data must go beyond the weakly
interacting theory and must include these effects in
combination.  A further problem is that in this dilute but
relatively clean system the disorder, as measured by $(k_f
l)^{-1}$, is large but $\tau$ is also large. Much of the
experimental data is taken in the ballistic regime (when $k_B
T \tau/\hbar \sim 1$) whereas the standard theories are
formulated in the diffusion limited regime when $k_B T
\tau/\hbar \ll 1$.

     A further {\it caveat}, appropriate to p-SiGe, should also
be noted. Spin-fluctuations play an important role in any
interaction effect theory: in p-SiGe the strong spin-orbit
coupling means the ``spins'' involved are not pure $S = \pm
1/2$ spin states, but rather a doublet of $M_J = \pm 3/2$
states. This is in contrast to situation, for example, in Si-
MOSFETs and p-GaAs where the spins are much less strongly
coupled to the orbital motion. 

     Renormalisation Group theories
\cite{abrahams99,finkelstein84,castellani84,castellani98},
which have already been alluded to, consider some of these
problems. They treat, self consistently, the modification of
interactions by disorder and predict a metallic-like decrease
of resistance with temperature and a low temperature saturation
at a finite value. This is qualitatively very similar to what
is observed experimentally. The issue of ballistic rather than
diffusive motion has recently been addressed by a new theory
put forward by Zala, Narozhny and Aleiner \cite{zala01a} (ZNA)
which treats coulomb interaction effects in both these regimes
(and also  the intermediate regime where there is a cross-
over). They find, in addition to the standard lnT dependence
(cf.eqn 4 and 11.) a linear term with a slope, and even  {\it
sign}, that depends directly on the strength of the spin
fluctuations. When only the Hartree part of the interaction is
included the standard temperature dependent screening result
(eqn.12) is recovered; when the Fock part is added the linear
slope can have either sign depending on the strength of the
spin fluctuations. The lnT term also changes sign and is in
fact given by eqn. 11. That is the theory reproduces the RG
result. Because the linear term persists to T=0 an explicit lnT
behaviour is only expected to emerge at very low temperatures. 

     In RG theories the renormalisation of the spin fluctuation
amplitude by disorder becomes important when {\it g}, the
dimensionless conductivity, approaches one. The ZNA theory is
only formulated for small disorder, ie {\it g} large, and does
not therefore reproduce the low temperature metallic
renormalisation predicted by the RG theories.

     In the ZNA theory the interaction strength is
characterised by the Fermi liquid parameter $F_0^{\sigma}$
describing the renormalisation of the spin susceptibility
[$\chi = \chi_0/(1+F_0^{\sigma})$] with a value of -1
corresponding to the ferromagnetic Stoner instability. It is
related to $\gamma_2$, used in RG theories, by
\cite{narozhnyprivate}

\begin{equation}
F_0^{\sigma}  =  - \gamma_2 / (1 + \gamma_2).
\label{eq14x}
\end{equation}

     In a subsequent paper Zala, Narozhny and Aleiner
\cite{zala01b} have also calculated the Hall coefficient. An
important conclusion here is that the ratio of two in eqn.8,
relating $\Delta R_H$ and $\Delta \sigma_{xx}^{ee}$, only holds
in the T=0 limit. As T increases it decreases and is typically
less than one when $k_BT\tau/\hbar$ exceeds about 0.1.

     The ZNA theory therefore advances on the ``old'' theories
in two important respects. It explicitly explains the high
temperature linear dependence of conductivity (and allows it to
be parametrised by a Fermi liquid parameter) and it provides a
quantitatively different interpretation for the temperature
dependence of the Hall data. It is of interest therefore to re-
examine the experimental data presented above using this
theory.

     Figure 10 compares  $\Delta \sigma_{xx}$, measured in
samples A and B, with the ZNA theory \cite{footnote2}. A weak
localisation term, $(e^2/\pi h) \ln(T)$, has been added to the
theoretical curves. In all cases there is an arbitary vertical
off-set, both experimentally (because $\sigma_0$ is not well
known) and theoretically (because of the ultra-violet
divergence in the lnT term). The data has therefore been
adjusted to provide a good match of the slopes.  Note also that
there is an experimental error of order 20\% in $\tau$
associated with the uncertainties in $\sigma_0$ and in the
effective mass. Values of $F_0^{\sigma}$ between about -0.55
and -0.65 (ie $\gamma_2$ between about 1.2 and 1.8) give a
satisfactory fit to the data. As shown in the insert, fits
where the weak localisation term has been excluded are somewhat
worse.

     Figure 11 shows a corresponding comparison for Hall data.
For both samples the values of  $F_0^{\sigma}$ needed to
explain the data (-0.3 to -0.5) are significantly smaller than
those obtained from the $\sigma_{xx}$ data.

     The values of $F_0^{\sigma}$ (and equivalently $\gamma_2$)
derived from the three separate experimental measurements are
summarised in Table I. They are of comparable magnitude,
corresponding to an enhancement of the spin susceptibility by
a factor of between 1.5 and 3 but are only approximately self-
consistent. It should be noted that for the Zeeman term at
least some of the measurements were made in the intermediate
temperature regime with values of $T\tau$ sufficiently large
that the RG eqns. 9 and 10 used to analyse the data probably
need to be replaced by a more general, intermediate
temperature, theory (not yet available). With this in mind the
values obtained for sample A are in quite reasonable agreement. 
For sample B, however, there is a disagreement by a factor of
order two between the values obtained from the temperature
dependence of $\sigma_{xx}$ and from the Hall coefficient. This
is not understood.

     Furthermore insight is obtained by fitting the ZNA theory
to the data for samples C and D in figure 9 where the absence
of a lnT term is establish down to quite low temperatures.
Better fits (see figure 12 ) are obtained because of the low
temperature cancellation between lnT terms of opposite sign
from the weak localisation and interaction effects. At higher
temperatures, the large linear slopes are also consistent with
values of $F_0^{\sigma}$ similar to those measured in samples
A and B. For sample C, however, there are still deviations from
the theory at low temperatures which indicate this might not be
the whole story and that the low temperature RG behaviour not
contained within the ZNA theory may also play a role. It should
also be noted that for samples B and C, which have very
similiar densities but mobilities differing by a factor two,
different values of $F_0^{\sigma}$ are required to fit the
linear depenendene. This implies that $F_0^{\sigma}$ is
determined not just by the density (ie by the value of $r_s$)
but also by the degree of disorder.

\subsection{Conclusions}

     Analysis of magneto-transport data in p-SiGe samples,
where the values of $r_s$ are significantly larger than one,
has allowed the separate identification of weak localisation
and Zeeman effect terms. Attempts to fit the results using the
standard theory for quantum corrections in weakly interacting
systems fail leads to inconsistencies between the fitting
parameters. There are, however, clear indications that
interaction effects are strong.  At B=0 the lnT dependence,
expected for both weak localisation and interaction effects, is
not observed. Rather there is the large linear increase of
resistance with increasing temperature that has been attributed
to the MIT transition. The saturation of this, at low
temperatures, is consistent with predictions of RG theories.

     Analysis of the data using the recent ``intermediate
temperature'' theories of Zala {\it et al} provides a much
better explanation of the data. Values of the spin triplet
interaction parameter, $F_0^{\sigma}$ of order -0.5 explain not
only the large magnitude of the Zeeman interaction term but
also the temperature dependence of the Hall coefficient and the
magnitude of the linear temperature dependence of the
resistivity. Deviations at the lowest temperatures suggest that
there might be a renormalisation of $F_0^{\sigma}$, as
predicted by Renormalisation Group theories but not including
in the ZNA theory.

     These results support the view that the MIT type of
behaviour observed in this system for clean samples is directly
associated with the large Coulomb interactions.

\subsection{Acknowledgements}

     H. Lafontaine and R.L. Williams are thanked for growth of
the samples, Y. Feng and J. Lapointe for sample
preparation and R. Dudek for technical assistance.
Helpful discussions with S. Kravchenko, B. Narozhny and S.
Studenikin are acknowledged.

\begin{table}

\caption
{Experimental values of $F_0^{\sigma}(\gamma_2)$. $\sigma(T)$
and Hall results are from fits to the ZNA theory; Zeeman
results from fits to eqns 6 and 10.}
\begin{tabular}{cccc}

Sample  &$\sigma(T)$         &Hall         &Zeeman    \\      
                                        \hline
A      &-0.55 (1.2)      &-0.45 (.85)       &-0.72 (2.6) \\ B 
     &-0.65 (1.9)     &-0.35 (.55)       &-0.38 (0.6)    
\end{tabular}
\label{table1}
\end{table}

\clearpage

\begin{figure} 
\caption  
{Magnetoresistance data for sample A at temperatures (shown)
between 0.15 and 2.2 K.   Lines are fitted to the weak
localisation peak (eqn.5) in the low field regime as described
in the text. 
}
\label{fig1}  
\end{figure} 
  
\begin{figure} 
\caption  
{Dephasing times,  $\tau_{\phi}$, extracted from fits to the
weak localisation peak in sample A (solid points) and sample B
(open points). The lines are the values given by eqn.13 with
f(g) = 7.4 (sample A, short dashes) and 2.6 (sample B, long
dashes). For comparison $\tau$, the transport lifetime is
approximately 0.9ps in both samples.
}
\label{fig2}  
\end{figure} 
  
\begin{figure} 
\caption  
{(a) Residues from the weak localisation fits to the low field
magnetoresistance for sample A. Temperatures are
respectively 250 (largest resistivity values),350, 450, 600,
750 and 1000 mK. (b) Data plotted against B/T. Solid line is a
fit to G(b) (eqn.6) and the dashed lines fits to eqns. 9 and
10. For b=1, B/T is approximately 0.4 tesla/kelvin.
}  
\label{fig3}  
\end{figure}  
  
\begin{figure} 
\caption  
{Magnetoresistance data for sample B at temperatures of 0.28
(lowest values), 0.37, 0.48, 0.65, 0.85 and 1.20K. At the
highest fields and lowest temperatures the Shubnikov-de Haas
oscillations are just becoming apparent.
}
\label{fig4}  
\end{figure}  
  
\begin{figure} 
\caption  
{Magnetic field dependence of the conductivity in sample B
derived from the data in figure 4. Dashed lines are fits to
eqn.5 plus a quadratic term for $B/T \leq$ 0.4 tesla/kelvin. 
}
\label{fig5}  
\end{figure}
  
\begin{figure} 
\caption  
{Residues from data in figure 5 with weak localisation(eqn.5)
and classical (eqn.7) terms subtracted off, plotted as a
function of B/T. Dashed line is fit to G(b) used to obtain a
value for $F^{\ast}$ and (in the quadratic region, with
eqn.10) $\gamma_2$. The smooth downturn in the data at the
highest fields reflects the onset of the Shubnikov-de Haas
oscillations.
}  
\label{fig6}  
\end{figure}  
  
\begin{figure} 
\caption  
{Hall coefficient as a function of field in sample A at
temperatures of 0.27, 0.39, 0.58, 0.85 and 1.20K ($R_H$
decreases with increasing temperature). The low field data,
dashed line, is a cubic fit to the $\rho_{xy}$ data matched to
direct measurements of $R_H$ at higher fields.  The
periodicity of the Shubnikov-de Haas oscillations corresponds
to a Hall coefficient of 1090 ohms/tesla.
}  
\label{fig7}  
\end{figure}  
  
\begin{figure} 
\caption  
{Hall coefficient as a function of field in sample B at
temperatures of 0.28, 0.37, 0.48, 0.65, 0.85 and 1.20K. Inset
shows the B=0 temperature dependence plotted against ln(T). The
low field data, dashed line, is a cubic fit to the
$\rho_{xy}$ data matched to direct measurements of $R_H$ at
higher fields.  The periodicity of the Shubnikov-de Haas
oscillations that are clearly visible corresponds to a Hall
coefficient of 5390 ohms/tesla.
}  
\label{fig8}  
\end{figure}  
    
\begin{figure} 
\caption  
{Zero field conductivity plotted against T/T$_F$ for: sample A
(T$_F$ = 76K),  sample B (T$_F$ = 16K), sample C (with a
similar density to sample B, T$_F$ = 17K, but a much higher
peak mobility of 21,000 cm$^2$(Vs)$^{-1}$) and (in the inset)
sample D (T$_F$ = 70K) with a density slightly less than sample
A . In each case the solid points are measured data and the
open points have a lnT weak localisation correction
subtracted. The solid lines are linear fits to the high T
values with $C_p$ = 2.5, 5.0, 6.1 and 2.5 respectively for
samples A - D.
}
\label{fig9}  
\end{figure}  

\begin{figure} 
\caption  
{Conductivity correction plotted against $T\tau$, where $\tau$
is the Drude lifetime, compared with the predictions of the ZNA
theory for various values of $F_0^{\sigma}$. A weak
localisation term has been added to the theoretical curves.
Open points, sample A; solid points, sample B. As noted in the
text there is an arbitary vertical off-set for both the
theoretical and experimental curves. The inset shows the same
comparison but without a weak localisation term included in the
theoretical curves.
}
\label{fig10} 
\end{figure}  

\begin{figure} 
\caption  
{Correction for the Hall data (around B=0) compared with the
predictions of the ZNA theory. Open points, sample A; solid
points, sample B. The Drude conductivity $\sigma_D$ is obtained
from experimental values in the low temperature limit.
}
\label{fig11} 
\end{figure}  

\begin{figure} 
\caption  
{As in figure 10 but for sample C (solid points) and
sample D (open points).
}  
\label{fig12} 
\end{figure}  


\begin{references}
\bibitem{footnote1} It is important to be sure that any field
or temperature dependence measured in the Hall coefficient is
not a spurious effect associated with feed-through of a small
amount of the $\rho_{xx}$ term. The zero of magnetic field can
be accurately determined and in all cases the offset in
$\rho_{xy}$ at B=0 is subtracted from the $\rho_{xy}$ data
before $R_H$ is calculated. In sample B, where feed-through
effects are most likely to be important, this off-set
corresponds to 0.4\% of $\rho_{xx}$ and has the expected
temperature dependence.  While subtracting this term eliminates
most of any feed-through there is the possibility a small B
dependent term may remain. Up to 0.2 tesla the maximum observed
magnetoresistance in $\rho_{xx}$ is about 2.5\% (at 0.28K)
corresponding to a contribution to the measured $\rho_{xy}$
values of about 1 ohm. This may be compared with the
experimentally observed deviation from linearity of about 85
ohms at 0.2 tesla. At lower fields any B dependent feed-through
was eliminated by making cubic fits to the data, over a field
range of -0.1 to +0.2 tesla, with the quadratic term set to
zero. That is the fit was forced to have the asymmetry required
of the Hall resistivity and reject any B dependent $\rho_{xx}$
contribution. A similar analysis for sample A shows even
smaller effects so for the field and temperature dependences of
$R_H$ reported in figures 7 and 8, only a very small
contribution, certainly less than 0.1\% of $R_H$, might be
attributed to spurious feed-through of a $\rho_{xx}$ term.
\bibitem{footnote2} The theoretical curves shown in figure 10
have been calculated assuming the cross-over functions $f(x)$
and $t(x,F_0^{\sigma})$ are 1. Over the range shown this will
lead to small errors, of order 0.1 e$^2$/h.
\end{references}
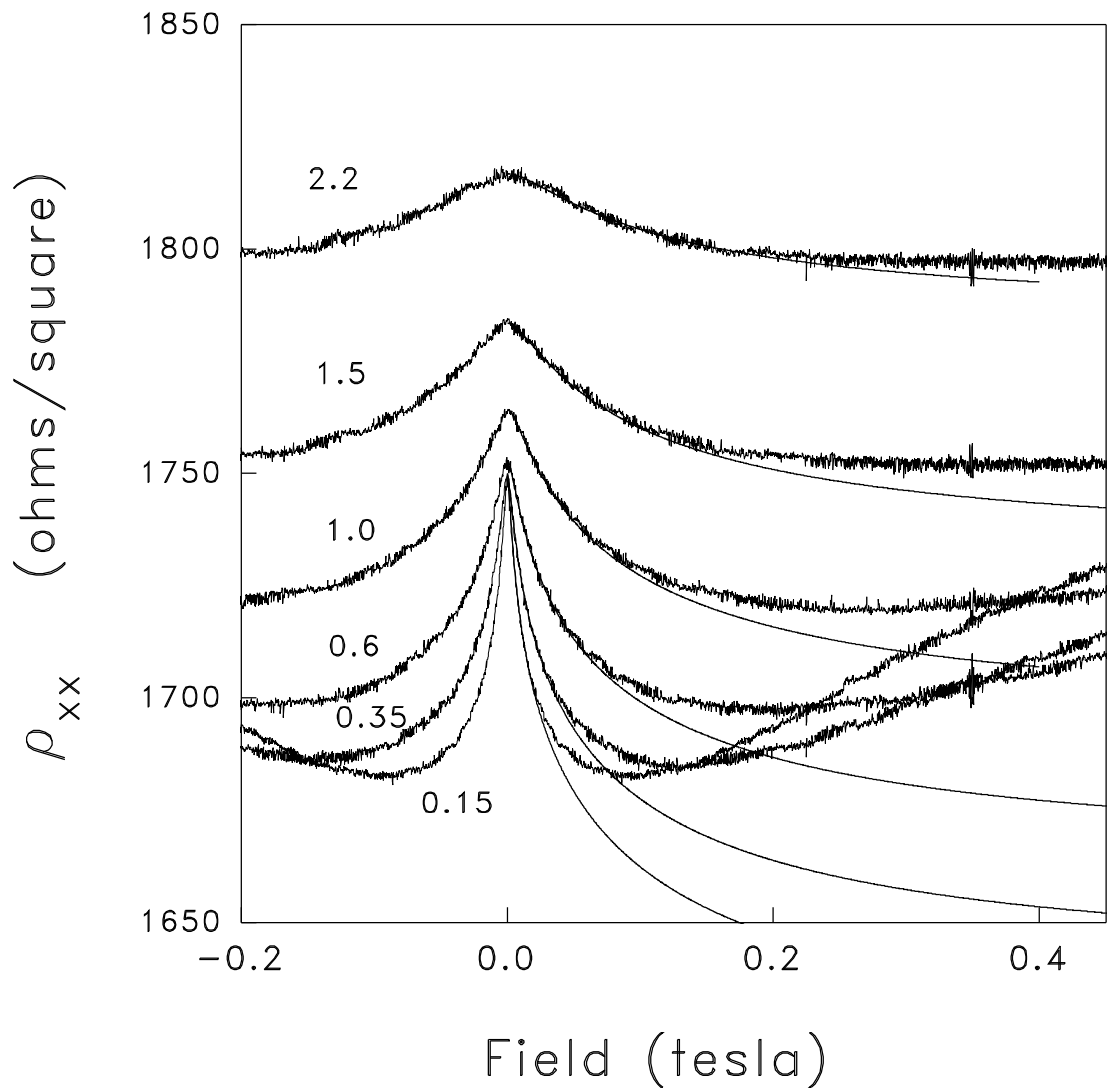
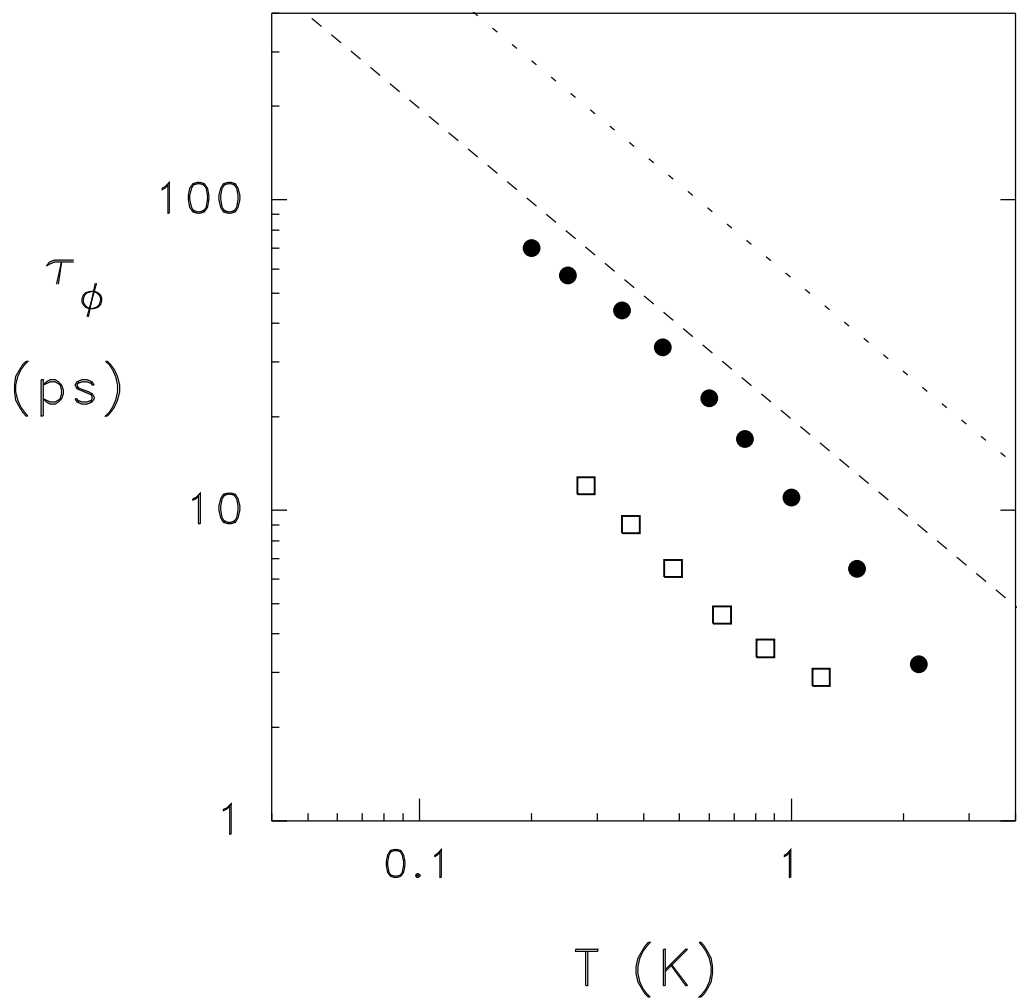
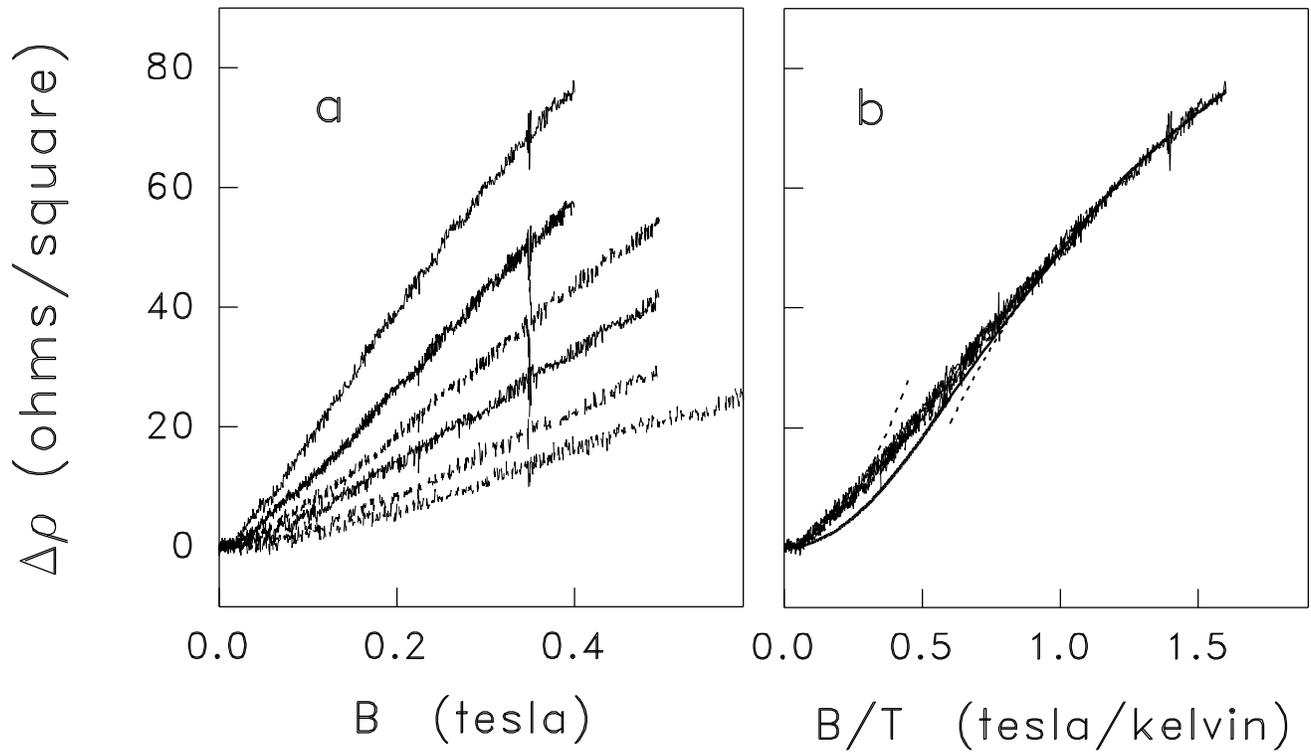
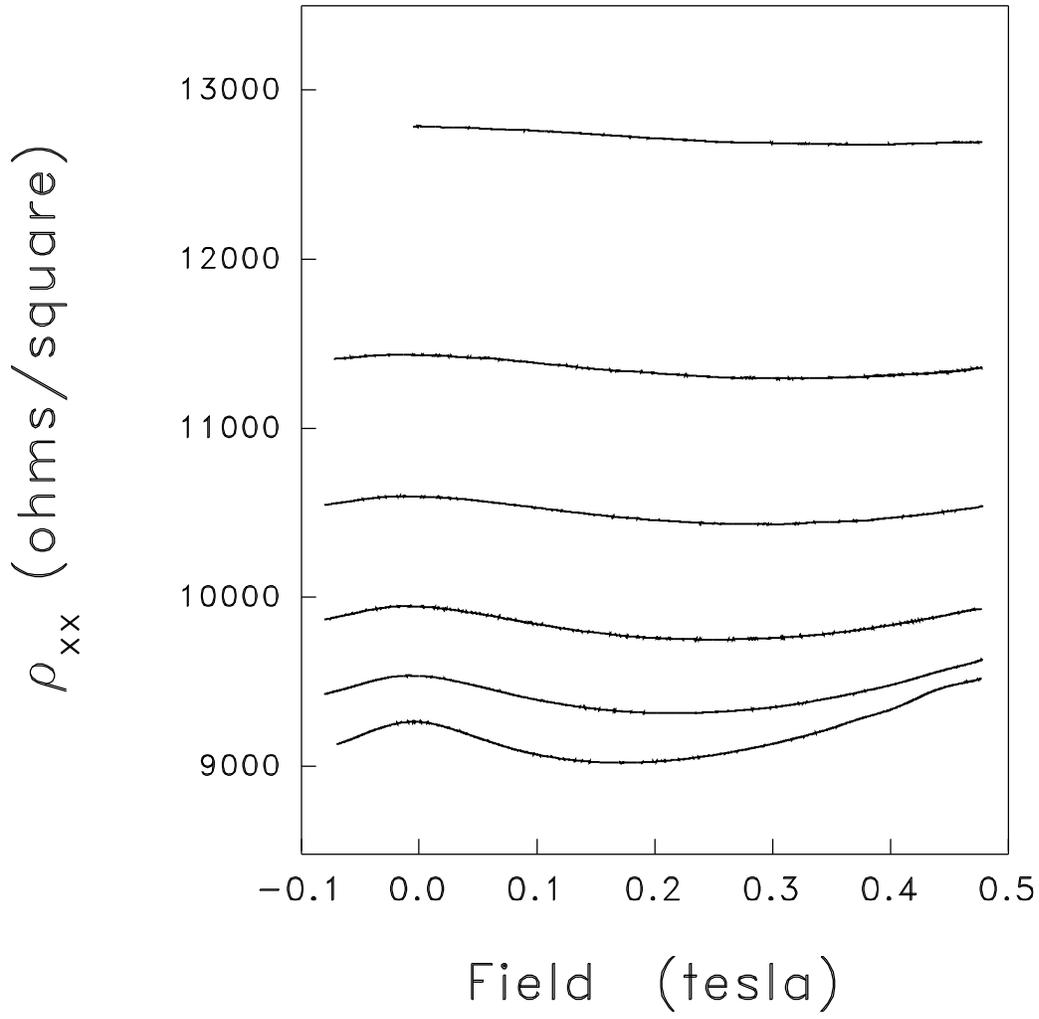
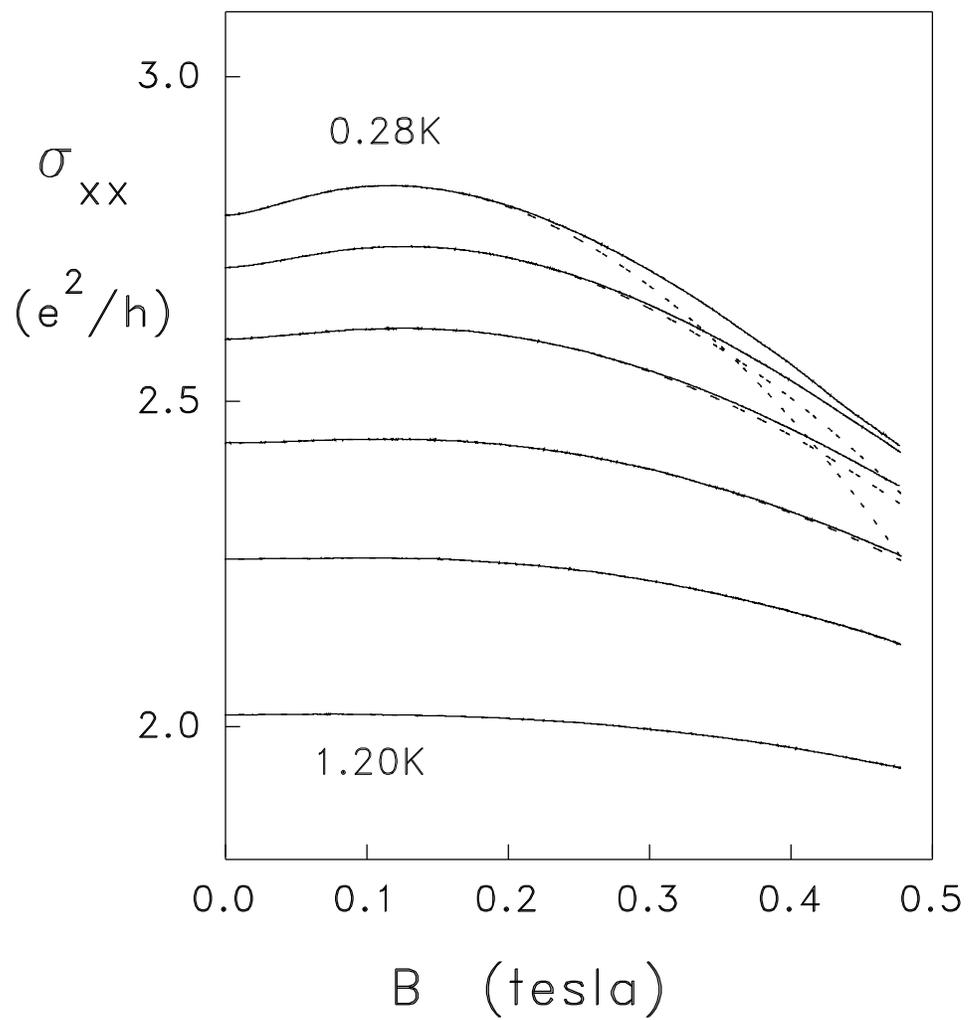
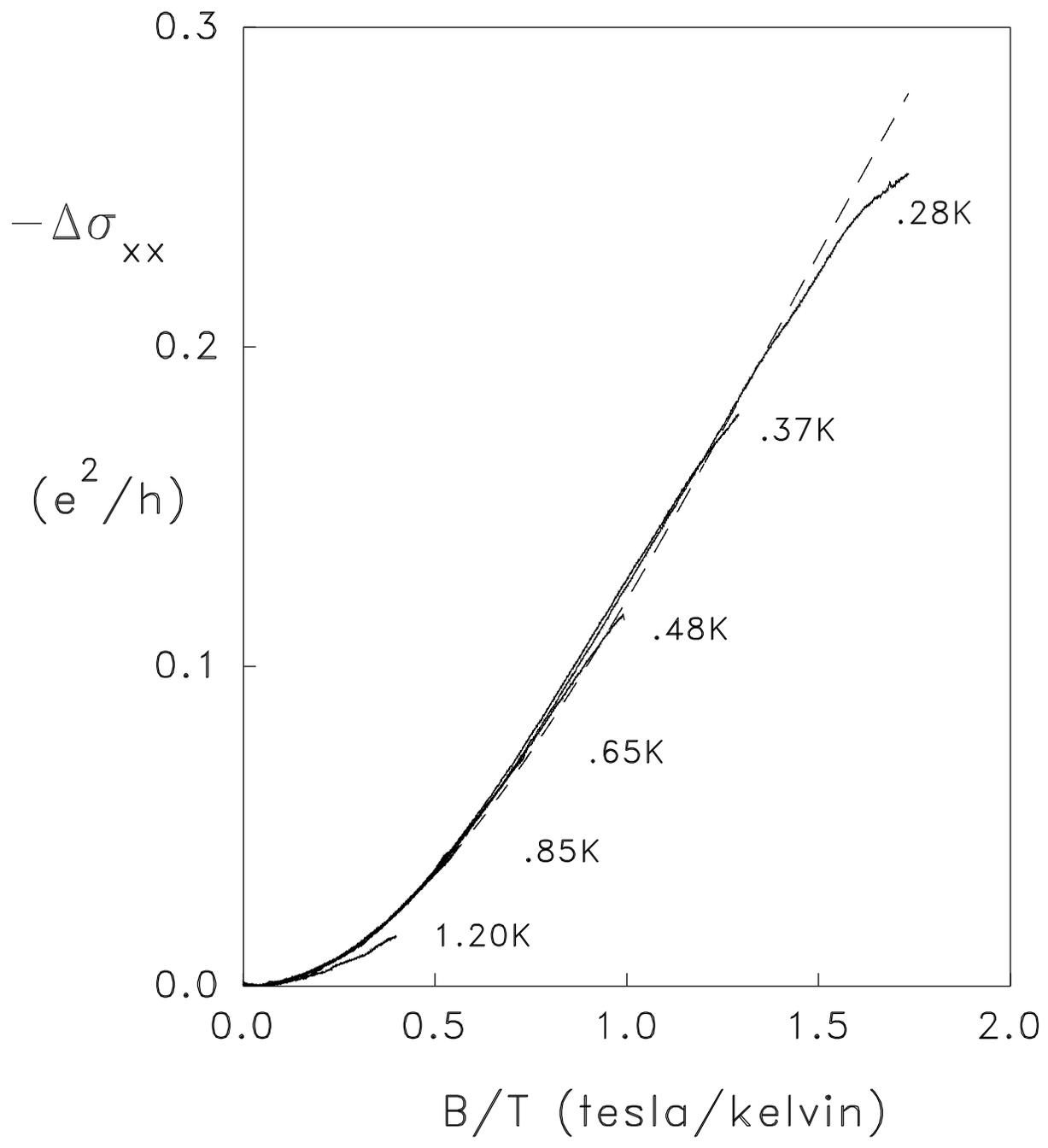
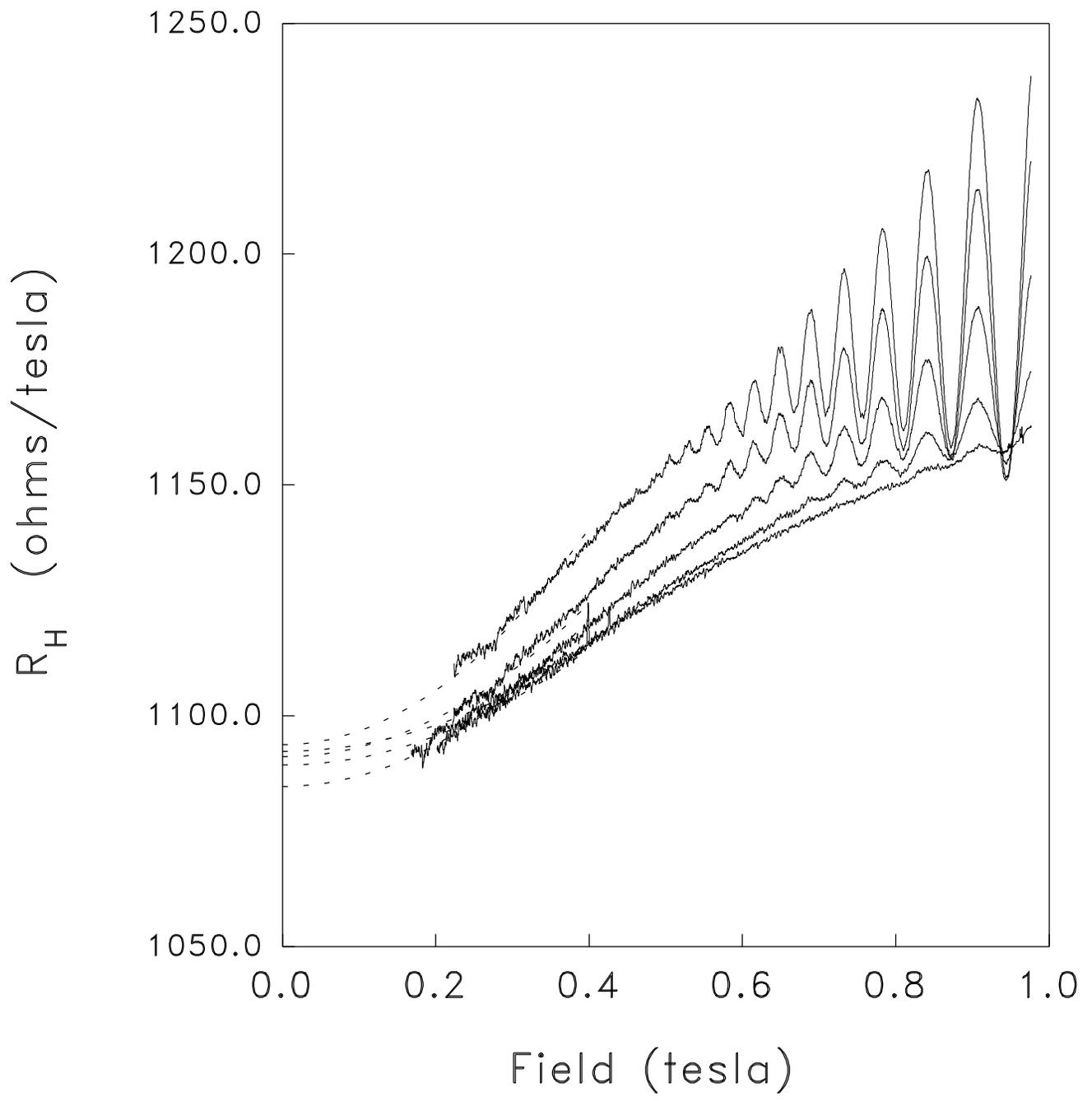
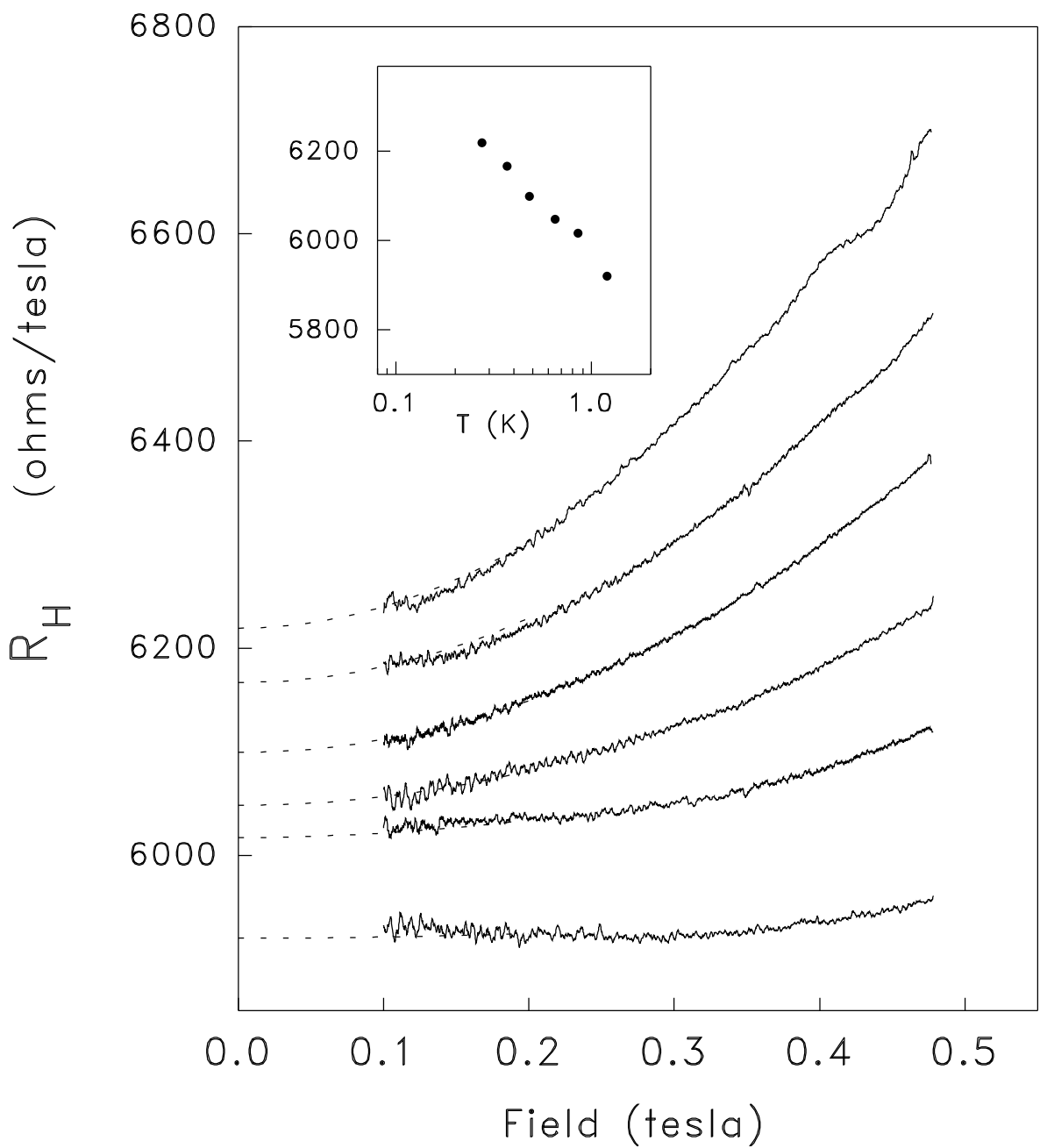
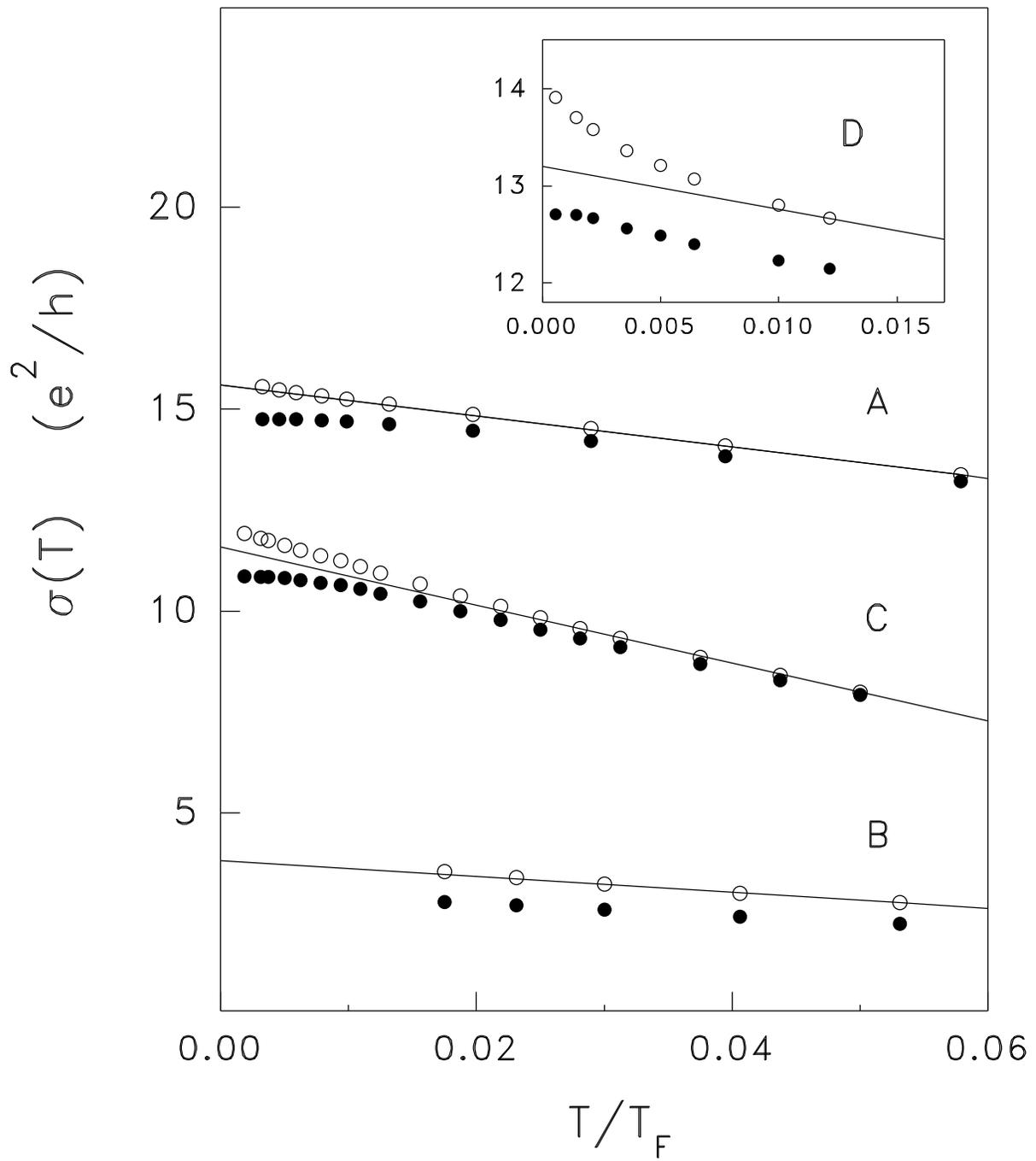
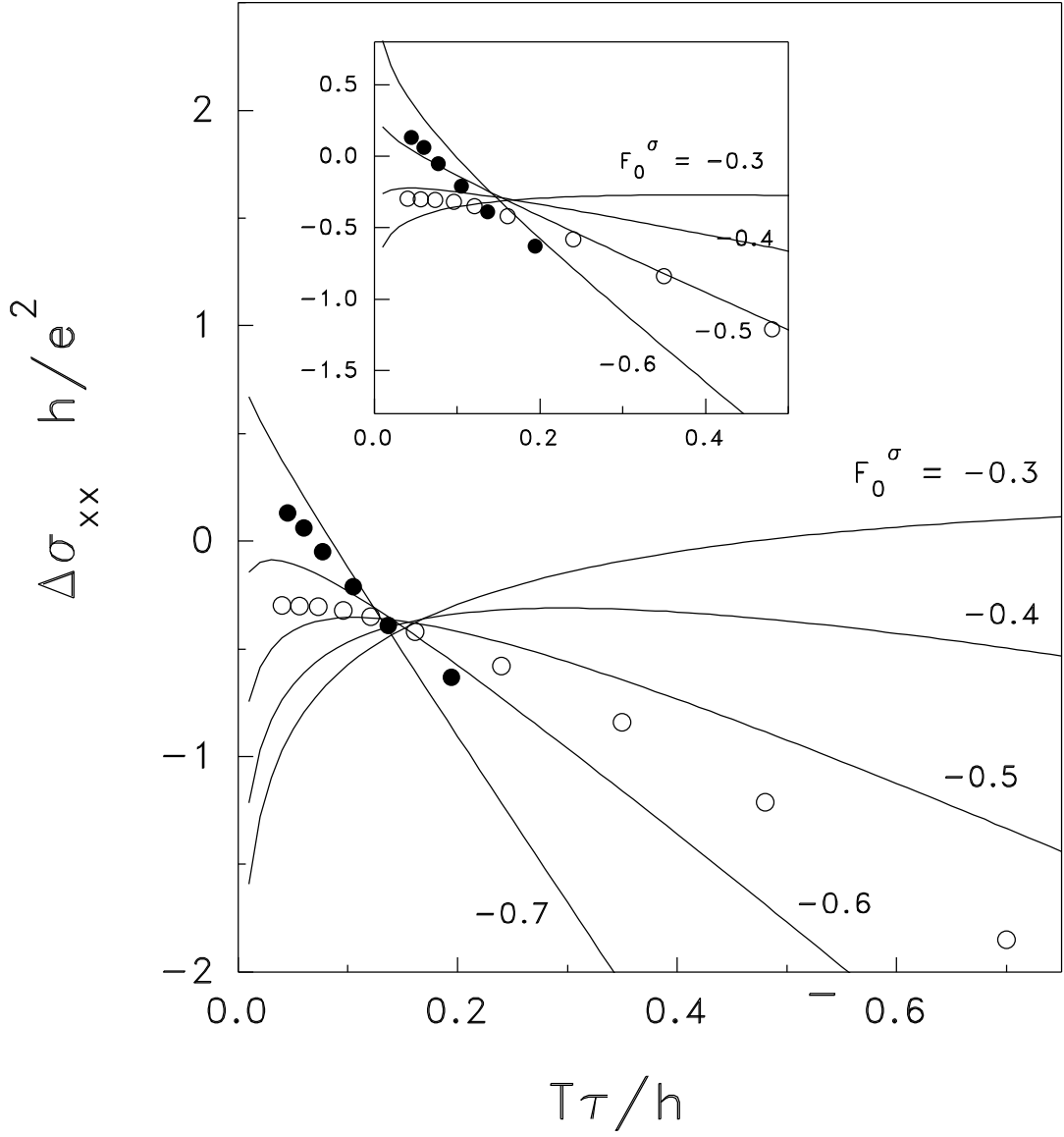
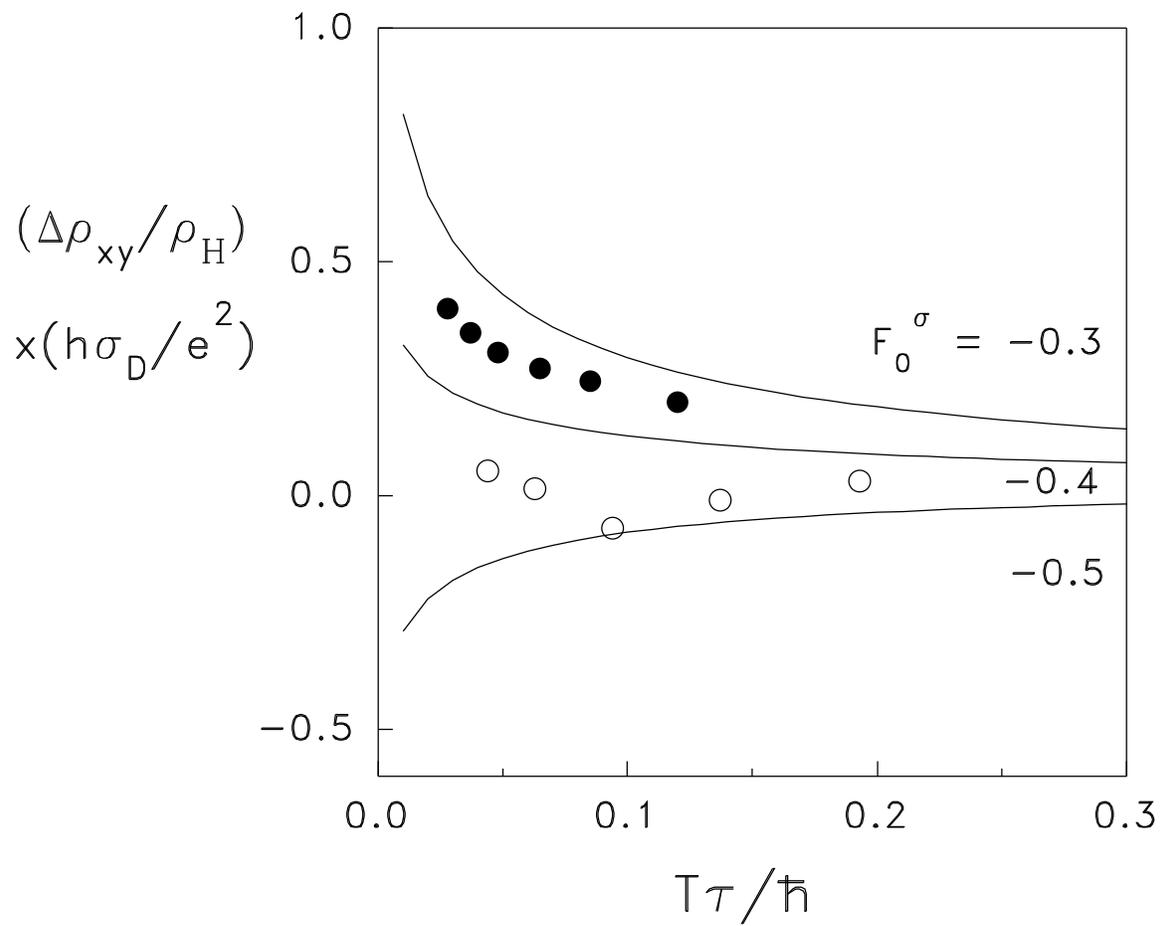
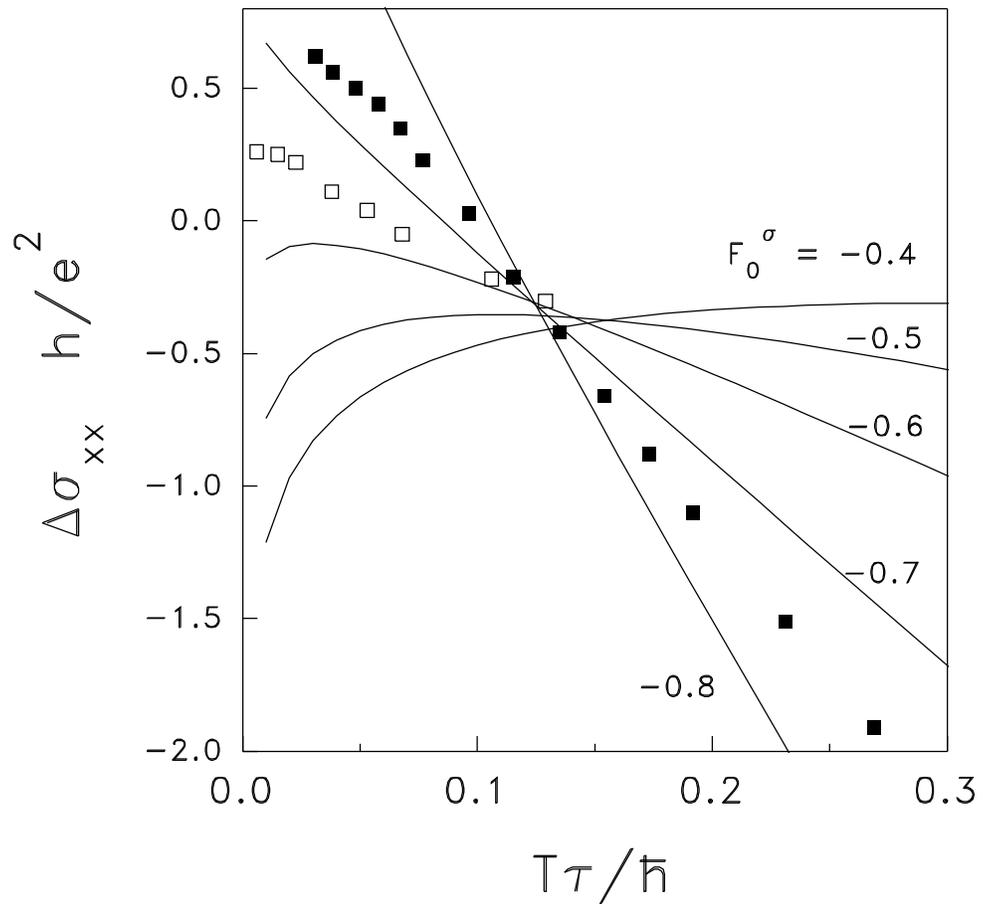
\end{document}